# Chiral magnetism in an altermagnet ($MnF_2$)

S. W. Lovesey [1,2,3]

[1]*ISIS Facility, STFC, Didcot, Oxfordshire OX11 0QX, United Kingdom*

[2]*Diamond Light Source, Harwell Science and Innovation Campus, Didcot, Oxfordshire OX11 0DE, United Kingdom*

[3] *Department of Physics, Oxford University, Oxford OX1 3PU, UK*

**Abstract** Symmetry informed diffraction patterns for magnetically ordered $MnF_2$ illuminated by x-rays tuned in energy to a Mn atomic resonance depend on circular polarization in the primary beam. The change in intensity of a Bragg spot with a change in handedness establishes a chiral signature for the fully compensated collinear antiferromagnet. The signature is calculated for electric dipole (E1) and electric quadrupole (E2) axial absorption events. Polar E1-E2 absorption events are forbidden by inversion symmetry in Wyckoff positions assigned to Mn ions in the established $MnF_2$ magnetic symmetry. It is compatible with atomic altermagnetism, and the corresponding order parameter contributes to E2-E2 and magnetic neutron Bragg diffraction patterns. Moreover, spin-flip patterns from polarized neutron diffraction depend on electronic quadrupole and octupole moments that are zero for the nominal $3d^5$ configuration of $Mn^{2+}$, which make them good tests of the actual electronic structure.

## 1. Introduction

Scattering techniques often have much to offer in gathering incisive information on properties of solids not available with other experimental methods. Distributions of magnetization in many crystalline materials have been measured on a sub-atomic scale and compared to state-of-the-art simulations (Bunău *et al*., 2022). If the periodicity of the magnetic and chemical structures are the same then magnetic and nuclear scattering of neutrons occur at the same points in reciprocal space and interfere with one another. In favourable cases the polarization dependence of the interference allows accurate determinations of the magnetic amplitude. The required interference is not available in magnetoelectric solids defined by p(parity)t(time)-symmetry (Landau *et al*., 1984), because magnetic and nuclear contributions to the neutron scattering amplitude are $90^o$ out of phase (Brown, 1993; Brown *et al*., 2002). In which case, magnetic and nuclear contributions to the intensity of a Bragg spot are in quadrature.

From a historical perspective it might be said that manganese fluoride ($MnF_2$; $3d^5$, $Mn^{2+}$) is to antiferromagnetism what eskolaite ($Cr_2O_3$) is to the magnetoelectric effect. Below an ordering temperature ≈ 67 K it is a tetragonal two-sublattice compensated antiferromagnet with Mn dipoles depicted in Fig. 1 parallel to the principal crystal axis (de Haas *et al*., 1940; Alperin *et al*., 1962; Yamani *et al*., 2010). An appreciation of altermagnetism in the past few years has propelled interest in $MnF_2$ and similar compounds (Yuan *et al*., 2020; Radaelli, 2024; Bhowal & Spaldin, 2024; Tamang *et al*., 2025). Altermagnets possess electronic bands with lifted spin degeneracy, which is a potentially useful attribute in the fabrication of next-generation devices. Our symmetry informed resonant x-ray Bragg diffraction patterns for $MnF_2$ demonstrate that its magnetic structure is chiral. Specifically, the intensity of a magnetic Bragg spot is influenced by circular polarization in the primary x-ray beam tuned in energy to a Mn atomic resonance. Notably, it is forbidden, along with the piezomagnetic effect, by magnetic symmetry ($\bar{1}'$) required for the linear magnetoelectric effect, e.g., $Cr_2O_3$ (Lovesey a, 2023). Furthermore, resonant x-ray and magnetic neutron diffraction unveil the altermagnetic order parameter for $MnF_2$.

An atomic resonance in the x-ray absorption spectrum is often a sharp feature (Collins & Bombardi, 2010; Paolasini, 2014). In which case, it is meaningful to assign an amplitude to the resonant contribution equal to its energy-integrated intensity. Four amplitudes are labelled by polarization states depicted in Fig. 2, and they can be developed in electronic multipoles introduced in Section 2 (Lovesey *et al.*, 2005; Lovesey & Balcar, 2013). Analytic expressions for axial multipoles for an informative atomic model are listed by Lovesey & Scagnoli (2009). In notation depicted in Fig. 2, ($\pi'\sigma$) denotes a rotated amplitude, and $|(\pi'\sigma)|^2$ the intensity of the Bragg spot enhanced by the atomic resonance. Universal expressions for diffraction amplitudes employed here are functions of the rotation of the illuminated crystal about the reflection vector by an angle $\psi$ (Scagnoli & Lovesey, 2009). Cell dimensions of $MnF_2$ and the energies of Mn atomic resonances mean that the Laue condition for Bragg diffraction is satisfied at the Mn K absorption edge and not at Mn L-edges; actual calculations of the condition are given in Section 3.

At room temperature haematite ($\alpha$-$Fe_2O_3$) and the Mott-Hubbard compound $V_2O_3$ possess the corundum structure. Resonant x-ray Bragg diffraction by antiferromagnetic $\alpha$-$Fe_2O_3$ exploiting the iron K edge (7.111 keV) is consistent with null electric dipole (E1) enhancement, and electric quadrupole (E2) azimuthal-angle scans (Finkelstein *et al.*, 1992). A peculiarity of enhancement at the K edge is that spin degrees of freedom in the valence state make no contribution to electronic multipoles (Lovesey, 1998). This means that magnetic multipoles for the ferric ion $Fe^{3+}$ ($3d^5$) available in diffraction at the K absorption edge are zero, because the nominally pure atomic state has zero orbital angular momentum. A detailed study of $\alpha$-$Fe_2O_3$ showed, beyond reasonable doubt, that magnetic multipoles contribute to the diffraction pattern, a result that implies unquenched orbital angular momentum in the valence state of the material (Rodríguez-Fernández *et al.*, 2013). An experimental study of monoclinic antiferromagnetic $V_2O_3$ using the vanadium K edge (5.465 keV) showed strong E2 enhancement in the rotated channel of polarization (Paolasini *et al.*, 2001). Bragg diffraction studies of orthorhombic solids $K_2CrO_4$ and $TbMnO_3$ revealed strong enhancements using a K edge absorption event (Cr K edge 5.994 keV; Fernández-Rodríguez *et al.*, 2008: Mn K edge 6.552 keV; Mannix *et al.*, 2007). Azimuthal angle scans for the magnetic perovskite are consistent with null diffraction in the unrotated channel ($\sigma'\sigma$) and strong diffraction in the rotated channel ($\pi'\sigma$) of polarization and E1 enhancement (Scagnoli & Lovesey, 2009).

## 2. Crystal and magnetic structures of $MnF_2$

Manganese ions occupy positions (0, 0, 0) and (1/2, 1/2, 1/2) in $P4_2/mnm$ (No. 136) that are centres of inversion symmetry. Cell dimensions are $a = b \approx 4.8736$ Å, $c \approx 3.3000$ Å (Yamani *et al.*, 2010). Fluoride $F^{1-}$ ions depicted in Fig. 1 are located in non-centrosymmetric positions between Mn ions.

For an atomic description of charge, orbital and spin degrees of freedom, Mn ions are assigned electronic spherical multipoles $\langle O^K_Q \rangle$ of integer rank K (Lovesey *et al.*, 2005; Lovesey & Balcar, 2013). Projections Q are in the range $- K \leq Q \leq + K$, and angular brackets denote an expectation, or time average, value of the enclosed quantum mechanical operator. In summary, multipoles encapsulate the electronic and magnetic ground state of Mn ions in $MnF_2$. Cartesian and spherical components $Q = 0, \pm 1$ of a vector $\mathbf{n} = (\xi, \eta, \zeta)$, for example, are related by $\xi = (n_{-1} - n_{+1})/\sqrt{2}$, $\eta = i(n_{-1} + n_{+1})/\sqrt{2}$, $\zeta = n_0$. A complex conjugate of a multipole is defined as $\langle O^K_Q \rangle^* = [(-1)^Q \langle O^K_{-Q} \rangle]$, and the diagonal multipole $\langle O^K_0 \rangle$ is purely real. The phase convention for real and imaginary parts labelled by single and double primes is $\langle O^K_Q \rangle = [\langle O^K_Q \rangle' + i\langle O^K_Q \rangle'']$. Whereupon $\langle O^1_\xi \rangle = -\sqrt{2} \langle O^1_{+1} \rangle'$ and $\langle O^1_\eta \rangle = -\sqrt{2} \langle O^1_{+1} \rangle''$.

The established symmetry of magnetically ordered $MnF_2$ (Fig. 1) is tetragonal $P4_2'/mnm'$ (No. 136.499, Belov-Neronova-Smirnova (BNS) setting of magnetic space groups, Bilbao Crystallographic server, http://www.cryst.ehu.es). The magnetic crystal class $4'/mmm'$ is centrosymmetric, permits a piezomagnetic effect, and forbids a ferromagnetic structure of axial magnetic dipoles. Wyckoff positions 2a possess oriented site symmetry $m.m'm'$ that includes inversion symmetry. In consequence, Mn multipoles are parity even (axial). Wyckoff position symmetry includes dyad operations $2_\zeta$ and $2'_{\xi\eta}$ that demand even projections Q, and $\langle O^K_Q \rangle = [(-1)^K \sigma_\theta \exp(i\pi Q/2) \langle O^K_{-Q} \rangle]$, where a time signature $\sigma_\theta = +1$ (non-magnetic, charge-like) or $\sigma_\theta = -1$ (magnetic). However, symmetry of the unit cell forbids some multipoles appearing in Bragg diffraction patterns.

Resonant x-ray diffraction can proceed with E1-E1 (K = 0-2) or E2-E2 (K = 0-4) absorption events, for which $\sigma_\theta = (-1)^K$ (Collins & Bombardi, 2010; Paolasini, 2014). Whereupon, the corresponding multipoles $\langle T^K_Q \rangle$ satisfy $\langle T^K_Q \rangle = [\exp(i\pi Q/2) \langle T^K_Q \rangle^*]$, and diagonal multipoles Q = 0 are allowed for all K, $\langle T^K_{+2} \rangle = i\langle T^K_{+2} \rangle''$ and $\langle T^K_{+4} \rangle = \langle T^K_{+4} \rangle'$. Axial magnetic multipoles in neutron diffraction $\langle t^K_Q \rangle$ are time-odd (Lovesey, 2015). They obey $\langle t^K_Q \rangle = [- (-1)^K \exp(i\pi Q/2) \langle t^K_Q \rangle^*]$ for $MnF_2$, meaning that $\langle t^K_0 \rangle$ can be different from zero for odd K. The atomic order parameter for altermagnetism $\langle O^3_{+2} \rangle$ is purely imaginary, and it is visible in both resonant x-ray and magnetic neutron Bragg diffraction patterns.

## 3. Resonant x-ray diffraction

Manganese absorption edges of immediate interest have energies $E \approx 6.552$ keV for the K edge (E1, $1s \rightarrow 4p$), and $L_2 \approx 0.649$ keV and $L_3 \approx 0.638$ keV (E1, $2p \rightarrow 3d$). The cell dimensions of $MnF_2$ are too short to satisfy the Laue condition for diffraction at manganese L edges. The Bragg angle in Fig. 2 is determined by $(\lambda/2a)$, where the photon wavelength $\lambda \approx (12.4/E)$ Å and the x-ray energy E is in units of keV. Whence, $(\lambda/2a) \approx 0.194$ for $MnF_2$ and the Mn K edge. The sensitivity to magnetic order at the $1s \rightarrow 4p$ dipole transition-energy is due to the 4p – 3d intra-atomic Coulomb interaction and to the mixing of the 4p with the 3d states of neighbouring magnetic ions. At the E2 threshold (E2, 1s $\rightarrow$ 3d) its origin is in the spin polarization of the 3d states.

A suitable electronic structure factor,

$$\Psi^{K}_{Q}(2a) = \sum \exp(i\, \boldsymbol{\kappa} \cdot \mathbf{d})\, \langle O^{K}_{Q}\rangle_{\mathbf{d}} = [\langle O^{K}_{Q}\rangle + (-1)^{h+k+l}(-1)^{K} \langle O^{K}_{Q}\rangle^{*}], \qquad (1)$$

where the sum is over the two positions **d** used by Mn ions, and the reflection vector $\boldsymbol{\kappa} = (h, k, l)$ with integer Miller indices. Multipoles are set in orthonormal axes ($\xi$, $\eta$, $\zeta$) that coincide with cell edges shown in Fig. 1. The result Eq. (1) follows from Wyckoff positions found in the Bilbao table MWYCKPOS for magnetic symmetry $P4_2'/mnm'$ (BNS). Wyckoff positions are related by operations listed in the table MGENPOS (BNS). Taken together, the two tables provide all information required to evaluate Eq. (1) and, thereafter, x-ray and neutron Bragg diffraction patterns. The reflection condition even $(h + k + l)$ for nuclear scattering follows from the requirement that $\Psi^{K}_{0}(2a)$ is different from zero for even K. Fluorine ions in Wyckoff positions 4f do not diffract at $(h, 0, l)$ with odd $(h + l)$, and these purely magnetic reflections are the subject of the present study of resonant x-ray Bragg diffraction. It follows that,

$$\Psi^{K}_{Q}(2a) = \langle T^{K}_{Q}\rangle\, [1 - (-1)^{K} \exp(i\pi Q/2)]. \qquad (2)$$

Specifically, $\Psi^{K}_{0}(2a) = 0$, $\Psi^{K}_{+2}(2a) = 2i \langle T^{K}_{+2}\rangle''$, $\Psi^{K}_{+4}(2a) = 0$ for even K, and $\Psi^{K}_{0}(2a) = 2\langle T^{K}_{0}\rangle$, $\Psi^{K}_{+2}(2a) = 0$, $\Psi^{K}_{+4}(2a) = 2 \langle T^{K}_{+4}\rangle'$ for odd K. Magnetic dipoles $\langle T^{1}_{0}\rangle$ are parallel to the c-crystal axis as in Fig. 1.

Diffraction amplitudes have been calculated with $\Psi^{K}_{Q}(2a)$ in Eq. (2) and universal expressions provided by Scagnoli & Lovesey (2009). Looking to individual amplitudes, Schmitt *et al.* (2021) list technical advantages in measuring the rotated amplitude $(\pi'\sigma)$. Using an E1-E1 absorption event, and notation $\cos(\chi) = (h/a)\, [(h/a)^2 + (l/c)^2\, ]^{-1/2}$ we find,

$$(\sigma'\sigma) = -\sin(\chi)\sin(2\psi)\, \langle T^{2}_{+2}\rangle'',$$

$$(\pi'\sigma) = -\cos(\chi)\cos(\theta)\sin(\psi)\, [(i/\surd 2)\, \langle T^{1}_{0}\rangle + \langle T^{2}_{+2}\rangle'']$$

$$+ \sin(\chi)\sin(\theta)\, [(i/\surd 2)\, \langle T^{1}_{0}\rangle - \cos(2\psi)\, \langle T^{2}_{+2}\rangle'']. \qquad (3)$$

Referring to Fig. 2, for the azimuthal angle $\psi = 0$ the $\eta$ and y axes are antiparallel. The quadrupole $\langle T^{2}_{+2}\rangle''$ contributes Templeton-Tempelton (T & T) scattering (Templeton & Tempelton, 1985). Notably, $(\sigma'\sigma)$ is purely real and $(\pi'\sigma)$ is complex. These features imply a non-zero chiral signature taken up in the next section.

## 4. Chiral signatures

Intensity of a Bragg spot picked out by circular polarization in the primary photon beam equals $P_2\Upsilon$ (Fernández-Rodríguez *et al.*, 2008) with,

$$\Upsilon = \{(\sigma'\pi)^*(\sigma'\sigma) + (\pi'\pi)^*(\pi'\sigma)\}'', \quad (4)$$

and a Stokes parameter $P_2$ (Landau *et al*., 1984; Lovesey *et al*., 2005) measures helicity in the primary x-ray beam. Intensity is a scalar quantity, and $\Upsilon$ and $P_2$ possess identical discrete symmetries, namely, both scalars are time-even and parity-odd. Partial intensity $P_2\Upsilon$ different from zero is a signature of diffraction by a chiral magnetic symmetry, of course. Evaluation of the chiral signature $\Upsilon$ demands a knowledge of all four diffraction amplitudes. For forbidden reflections $(2n + 1, 0, 0)$ and an E1-E1 absorption event,

$$(\sigma'\sigma) = 0,\ (\pi'\pi) = i \sin(2\theta) \cos(\psi) \langle T^1{}_0\rangle,$$

$$(\pi'\sigma) = -\cos(\theta) \sin(\psi) [(i/\surd 2) \langle T^1{}_0\rangle + \langle T^2{}_{+2}\rangle'],$$

$$\Upsilon = \sin(2\theta) \cos(\theta) \sin(2\psi) \langle T^1{}_0\rangle \langle T^2{}_{+2}\rangle''. \quad (5)$$

The crystal c axis is normal to the plane of scattering for $\psi = 0$. The chiral signature is created by inference between a magnetic dipole and T & T scattering. However, this is not a general result, e.g., the chiral signature for a Sohncke-type magnetic material $Pb(TiO)Cu_4(PO_4)_4$ includes a product of dipoles (Lovesey, 2024). Notably, there is no E1-E1 scattering in the unrotated channel $(\sigma'\sigma)$, which simplifies the calculation of the chiral signature. Results here for $(\sigma'\sigma)$ and $(\pi'\sigma)$ follow from Eq. (3) on setting the Miller index $l = 0$, and $\sin(\chi) = 0$. Amplitudes for an E2-E2 event are,

$$(\sigma'\sigma) = i \sin(2\theta) \cos(\psi) [-\langle T^1{}_0\rangle + \{5 \cos^2(\psi) - 3\} \langle T^3{}_0\rangle], \quad (6)$$

$$(\pi'\sigma) = i \sin(\psi) [\cos(3\theta) \langle T^1{}_0\rangle + \{3 \cos(3\theta) - 5 \cos(\theta) (3 \cos^2(\theta) - 2) \sin^2(\psi)\} \langle T^3{}_0\rangle$$
$$- i \surd(30/7)\ Z],$$

with a purely real sum of non-magnetic (T & T) multipoles,

$$Z = [\cos(3\theta) \langle T^2{}_{+2}\rangle'' + (1/\surd 3) \cos(\theta) \{(7 \cos^2(\psi) - 1) \cos^2(\theta) - 1\} \langle T^4{}_{+2}\rangle'].$$

Leading to a chiral signature,

$$\Upsilon = \sin(2\theta) \sin(2\psi)\ Z\ [\{8 \cos^2(\theta) - 5\} \langle T^1{}_0\rangle + \cos^2(\theta) \{5 \cos^2(\psi) - 3\} \langle T^3{}_0\rangle]. \quad (7)$$

Both chiral signatures are odd with respect to the azimuthal angle, and functions of $(2\psi)$. Manganese ions will possess unquenched orbital angular momentum if $\langle T^3{}_0\rangle$ is different from zero at the K edge, which has been established for ferric ions in haematite (Rodríguez-Fernández *et al*., 2013).

## 5. Neutron polarization analysis

Elastic and inelastic scattering of polarized neutrons from haematite and $MnF_2$ are included by Moon *et al*. (1969), and neutron Bragg diffraction by ordered $MnF_2$ is covered in depth by Alperin *et al*. (1962) and Yamani *et al*. (2010). Brown (1993) and Bourges *et al*. (2021) include additional relevant examples of polarized neutron diffraction by magnetic materials, while Lovesey & Watson (1998) introduce a general theoretical framework for elastic and inelastic scattering of polarized neutrons.

Unlike multipoles for resonant x-ray diffraction, multipoles in neutron diffraction are strong functions of the magnitude of the reflection vector. This feature is illustrated in a useful approximation to the axial dipole $\langle \mathbf{t}^1 \rangle$,

$$\langle \mathbf{t}^1 \rangle \approx (\langle \boldsymbol{\mu} \rangle/3)\, [\langle j_0(\kappa) \rangle + \langle j_2(\kappa) \rangle\, (g_o - 2)/\, g_o]. \qquad (8)$$

The quantities $\langle j_0(\kappa) \rangle$ and $\langle j_2(\kappa) \rangle$ are spherical Bessel functions averaged over the radial distribution of electrons in the valence state. In the forward direction of scattering $\langle j_0(0) \rangle = 1$ and $\langle j_2(0) \rangle = 0$, and the principal maximum of $\langle j_2(\kappa) \rangle$ occurs at $\kappa \approx 6.3$ Å$^{-1}$ (Fig. 3 in Lovesey b, 2023). Returning to Eq. (8), the magnetic moment $\langle \boldsymbol{\mu} \rangle = g_o \langle \mathbf{S} \rangle$ and the orbital moment $\langle \mathbf{L} \rangle = [(g_o - 2) \langle \mathbf{S} \rangle]$. The coefficient of $\langle \mathbf{L} \rangle$ is approximate, while $\langle \mathbf{t}^1 \rangle = (1/3) \langle 2\mathbf{S} + \mathbf{L} \rangle$ for $\kappa \rightarrow 0$ is an exact result. Higher order multipoles with even rank depend on the electronic position operator $\mathbf{n}$. The equivalent operator $[\langle j_2(\kappa) \rangle (\mathbf{S} \times \mathbf{n})\, \mathbf{n}]$ for $\mathbf{t}^2$ shows that the quadrupole measures the correlation between the spin anapole $(\mathbf{S} \times \mathbf{n})$ and orbital degrees of freedom (Lovesey, 2015). Magnetic neutron multipoles with an even rank do not exist for magnetic states derived from a state specified by one value of the total angular momentum J. Instead, a ground state must possess two, or more, J-states before $\langle \mathbf{t}^2 \rangle$ is non-zero (Lovesey, 2015). A single J-state is at odds with the basic premise of altermagnetism, because the singular state is one outcome of strong spin-orbit coupling (Tamang *et al*., 2025). Moreover, higher order multipoles are zero for the nominal atomic state $Mn^{2+}$ $(3d^5)$ (Lovesey, 2015). Evidence to the contrary in $MnF_2$ can be obtained from Bragg spots at overlapping nuclear-magnetic reflections indexed by even $(h + k + l)$.

Polarization analysis offers great sensitivity to magnetic contributions in Bragg spots of mixed origin. A polarization dependence of the neutron scattering can be described as a departure from unity of the ratio of the reflected intensities for incoming neutron beams of opposite (±) polarization, i.e., $(A_n + A_m)^2 /(A_n - A_m)^2$, where $A_n$ and $A_m$ are total nuclear and magnetic amplitudes, respectively. For an amplitude ratio that departs from unity neither $A_n$ nor $A_m$ can be zero. Moreover, $A_n$ and $A_m$ must have like phases (Brown, 1993). More generally, intensity of a magnetic Bragg spot = $|\langle \mathbf{Q}_\perp \rangle|^2$, where $\langle \mathbf{Q}_\perp \rangle = [\mathbf{e} \times (\langle \mathbf{Q} \rangle \times \mathbf{e})]$, the unit vector $\mathbf{e} = \boldsymbol{\kappa}/\kappa$, and the neutron scattering amplitude $\langle \mathbf{Q} \rangle$ is a sum of spin and orbital magnetizations illustrated with the dipole approximation in Eq. (8) (Lovesey, 2015). A fraction $\propto \{(1/2)\,(1 + P^2)\, |\langle \mathbf{Q}_\perp \rangle|^2 - |\mathbf{P} \cdot \langle \mathbf{Q}_\perp \rangle|^2\}$ of neutrons participate in events that change (flip) the neutron spin orientation, where $\mathbf{P}$ is the primary polarization. An assumption of perfect polarization $(\mathbf{P} \cdot \mathbf{P}) = 1$ yields a standard spin-flip signal (Moon *et al*., 1969; Bourges *et al*., 2021),

$$SF = [|\langle \mathbf{Q}_\perp \rangle|^2 - |\mathbf{P} \cdot \langle \mathbf{Q}_\perp \rangle|^2]. \qquad (9)$$

Evidently, all scattering is spin-flip when $\mathbf{P}$ and $\mathbf{e}$ are aligned since $\mathbf{e} \cdot \langle \mathbf{Q}_\perp \rangle = 0$. Amplitudes for $MnF_2$ correct to the level of magnetic octupoles, and even $(h + k + l)$ are (Lovesey, 2015),

$$\langle Q_\xi \rangle \approx (e_\eta\, e_\zeta) f,\ \ \langle Q_\eta \rangle \approx (e_\zeta\, e_\xi)\, f,\ \langle Q_\zeta \rangle \approx (e_\xi\, e_\eta)\, g, \qquad (10)$$

$$f = \langle t^{2}_{+2} \rangle' + (1/2)\sqrt{(35/2)}\, \langle t^{3}_{+2} \rangle'',\ (2f + g) = (3/2)\sqrt{(35/2)}\, \langle t^{3}_{+2} \rangle''.$$

The octupole $\langle t^{3}_{+2} \rangle''$ measures a bulk magnetic octupole in $MnF_2$. As such, it is a signature of altermagnetism that we recount in Section 6. On choosing $e_\eta = 0$ we note $2(e_\zeta\, e_\xi) = \sin(2\chi)$, where $\chi$ is an angle in Eq. (3). Standard choices for the neutron polarization include (a) **P** and **e** parallel (b) $\mathbf{e} \cdot \mathbf{P} = 0$ using $\mathbf{P} = (-\ e_\zeta,\ 0,\ e_\xi)$, and (c) $\mathbf{e} \cdot \mathbf{P} = 0$ using $\mathbf{P} = (0, 1, 0)$. The corresponding spin-flip signals are SF(a) = $[(e_\zeta\, e_\xi)\ f]^2$, SF(b) = SF(a), and SF(c) = 0. Thus, diffraction conditions (a) and (b) access $\langle t^{3}_{+2} \rangle''$.

## 6. Atomic altermagnetism

Spin states in the electronic bands of an altermagnet are not degenerate - they are spin split - even though bulk magnetization (ferromagnetism) is forbidden by symmetry. Conditions for this desirable property of an antiferromagnet are non-relativistic with no place for a spin-orbit coupling (Šmejkal *et al*., 2022; Tamang *et al*., 2025). The well-established Rashba-Dresselhaus effect removes spin degeneracy under distinctly different conditions (Dresselhaus, 1955; Rashba & Sheka, 1959; Manchon *et al*., 2015). It applies to nonmagnetic materials with finite spin-orbit coupling, and the absence of a centre of inversion symmetry, e.g., an acentric crystal structure. By elaborating a theoretical concept put forward by Peka & Rashba (1965), Yuan *et al*. (2020) demonstrate a spin-splitting that depends on the electron band momentum, and it is operates even in centrosymmetric antiferromagnets. Most of the 69 altermagnetic crystal classes permit the piezomagnetic effect (permitted in 66 of the total 90 magnetic crystal classes), and they include collinear antiferromagnets (Radaelli, 2024). Crucially, ferro-type ordering of magnetic axial multipoles with rank $K > 1$ is permitted, even though axial magnetic dipoles ($K = 1$) form a fully compensated antiferromagnetic structure. Other potentially useful material properties, all accomplished without ferromagnetism, are efficient spin-current generation, spin-splitting torque, giant magnetoresistance, and an anomalous Hall effect.

Manganese fluoride is an atomic altermagnetic with ferroically ordered axial magnetic octupoles, i.e., bulk octupole magnetism. They are denoted here by $\langle \mathbf{T}^3 \rangle$ and $\langle \mathbf{t}^3 \rangle$ in resonant x-ray and magnetic neutron diffraction, respectively. Our electronic structure factor Eq. (1) with Miller indices $h = k = l = 0$ delineates bulk properties of symmetry $P4_2'/mnm'$ (No. 136.499), specifically, $\Psi^{3}_{Q}(2a) = 2i\ \langle t^{3}_{Q} \rangle''$ for neutrons. In the application of symmetry for Wyckoff positions 2a we use $\sigma_\theta = -1$, which is correct for magnetic neutron diffraction and odd rank $\langle \mathbf{T}^{K} \rangle$. Continuing with a calculation of $\langle t^{3}_{Q} \rangle''$, position symmetry shows that it can be different from zero for projections $Q = \pm 2$. An E2-E2 absorption event accesses the altermagnetic order parameter $\langle T^{3}_{+2} \rangle''$. It is invisible in purely magnetic reflections, however. For space group allowed reflections the rotated amplitude $(\pi'\sigma)$ is preferred for technical reasons (Schmitt *et al*., 2021) and using $(2n, 0, 0)$,

$$(\pi'\sigma) = \sin(2\psi)\ [\sin(3\theta)\ \langle T^{2}_{0} \rangle + \sin(\theta)\ \{[(\sqrt{5}/3)\ (4 \cos^{2}(\theta) - 1 - 7\ (\cos(\theta) \sin(\psi))^{2})]\ \langle T^{4}_{0} \rangle$$

$$+ (\sqrt{14}/3)\ [1 - (\cos(\theta)\ \sin(\psi))^2]\ \langle T^4_{+4}\rangle'\} - i\sqrt{(7/6)}\ \{\sin(\theta) + \sin(3\theta)\}\langle T^3_{+2}\rangle''\}]. \quad (11)$$

Order parameter and T & T contributions to the E2-E2 amplitude are 90° out of phase, and they are in quadrature in the Bragg spot intensity $|(\pi'\sigma)|^2$. The phase shift between magnetic $\langle T^3_{+2}\rangle''$ and nonmagnetic T & T contributions we find in the $MnF_2$ amplitude for diffraction at a space group allowed reflection is the analogue of the the phase shift between magnetic and nuclear contributions to the neutron diffraction amplitude (Alperin *et al*., 1962; Brown 1993). Intensity is a four-fold periodic function of the azimuthal angle. At the origin of a scan in $\psi$ the crystal c axis is normal to the plane of scattering in Fig. 2.

**7. Conclusions**

In summary, we demonstrate that the established magnetic symmetry of $MnF_2$ is chiral. To this end, we made a symmetry informed calculation of magnetic Bragg spots including circular polarization in a primary beam of x-rays tuned in energy to a manganese atomic resonance. Specifically, non-magnetic (time-even and charge like) and magnetic (time-odd) contributions to the four diffraction amplitudes, labelled by four states of photon polarization in Fig. 2, are not of one phase. This finding is correct for axial electric dipole (E1-E1) and electric quadrupole (E2-E2) diffraction amplitudes. Polar E1-E2 amplitudes are forbidden by inversion symmetry in the Wyckoff positions used by Mn ions. By contrast, parity-time symmetry in the linear magnetoelectric effect imposes identical phases on time-even and time-odd contributions to the four amplitudes, and the intensity of a Bragg spot is immune to circular polarization in the primary beam of x-rays. Our scattering amplitudes and chiral signatures are functions of an azimuthal angle that measures rotation of the $MnF_2$ crystal about the reflection vector.

A study of polarization analysis of magnetic neutron diffraction by $MnF_2$ complements results for resonant x-ray Bragg diffraction. The emphasis is on magnetic quadrupoles and octupoles that contribute to mixed nuclear-magnetic Bragg spots. Since they are zero for a nominal electronic configuration $Mn^{2+}$ ($3d^5$) there is potential in future experiments to reveal subtle effects in the electronic structure.

The altermagnetic order parameter is exposed in both magnetic neutron and resonant x-ray Bragg diffraction patterns. In the latter case it appears in space group allowed Bragg spots with magnetic and nonmagnetic contributions 90° out of phase, which is anticipated from earlier studies of magnetic neutron diffraction by $MnF_2$.

**Acknowledgements** Professor P. G. Radaelli made his paper available before publication (Radaelli, 2024). Dr D. D. Khalyavin contributed to Section 6.

.

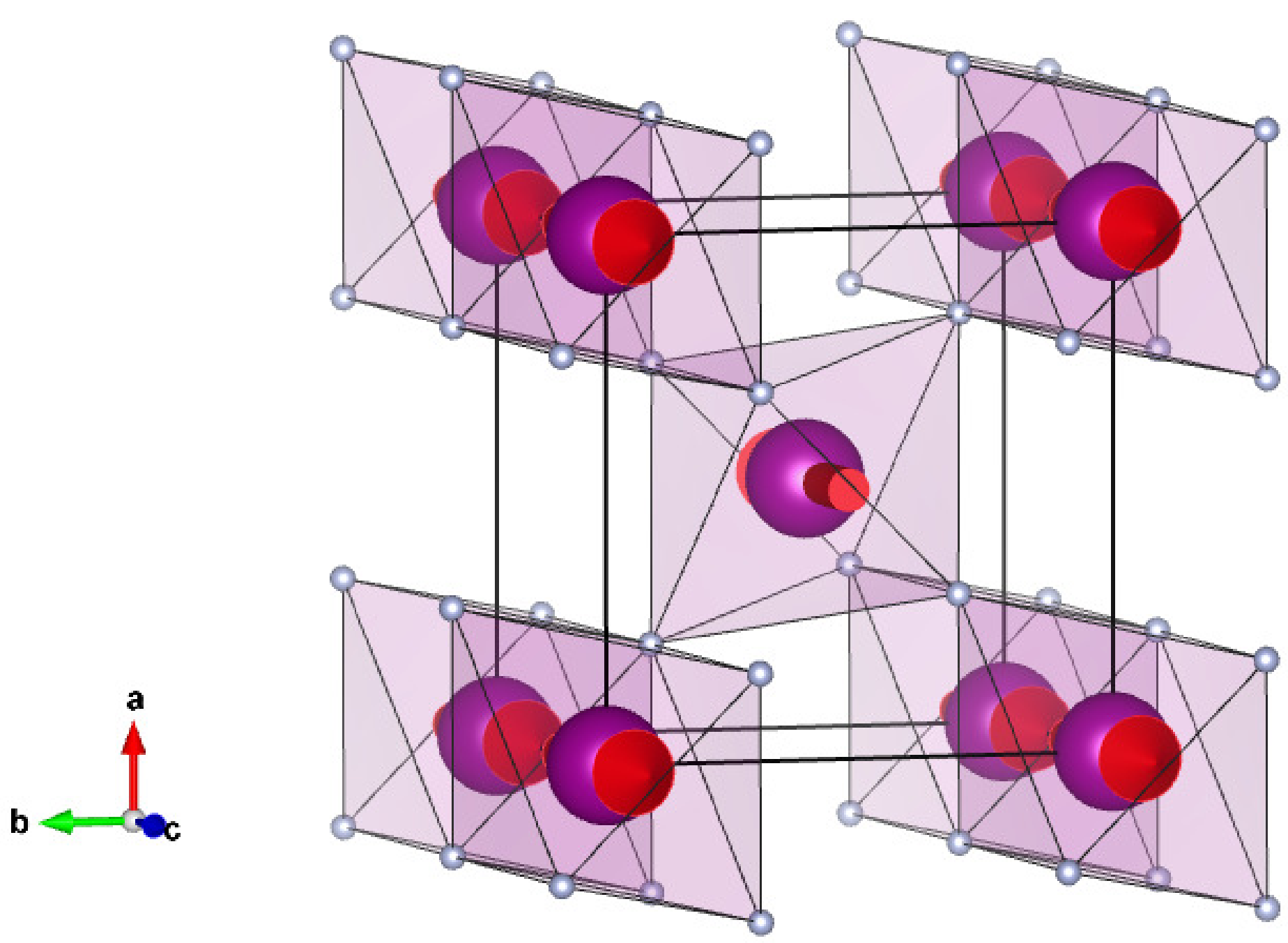

**Figure 1** Antiferromagnetic structure of $MnF_2$ including fluorine ions (reproduced from MAGNDATA, http://webbdcrista1.ehu.es/magndata). A transition metal ion is surrounded by six $F^{1-}$ ligands forming octahedra.

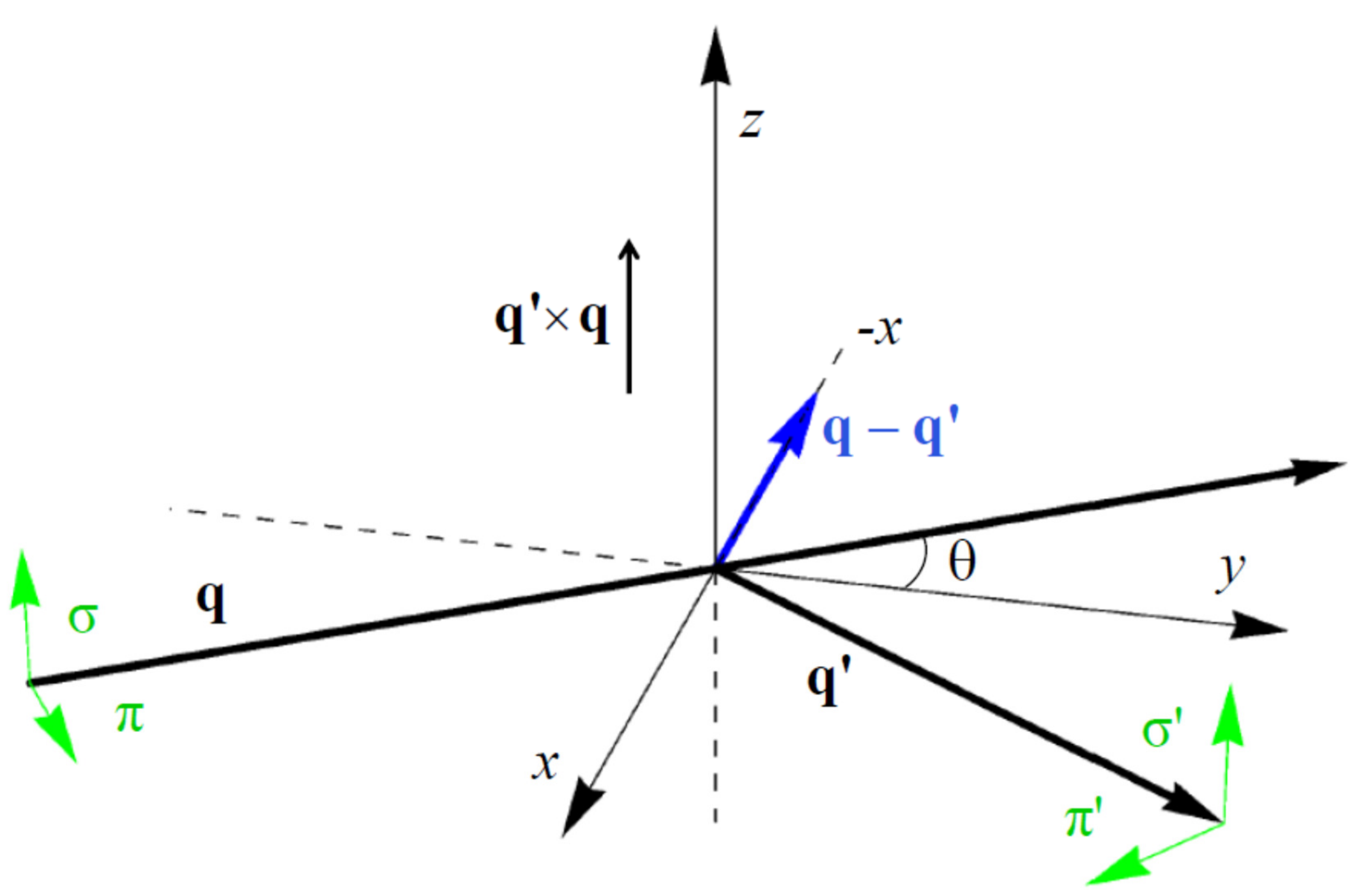

**Figure 2** Primary (σ, π) and secondary (σ′, π′) states of polarization. Corresponding wavevectors **q** and **q**′ subtend an angle 2θ. The Bragg condition for diffraction is met when **q** − **q**′ coincides with a reflection vector ($h$, $k$, $l$) of the reciprocal lattice. Crystal vectors that define local axes (ξ, η, ζ) and the depicted Cartesian (x, y, z) coincide in the nominal setting of the crystal.